\documentclass[superscriptaddress,aps,preprintnumbers,amsmath,showpacs,amssymb,prd,nofootinbib,reprint]{revtex4-1}

\usepackage[dvipdfmx]{graphicx}
\usepackage{amsfonts}
\usepackage[mathscr]{eucal}
\usepackage{ascmac}
\usepackage{dcolumn}% Align table columns on decimal point
\usepackage{bm}% bold math
\usepackage[colorlinks=true,linkcolor=blue,citecolor=blue]{hyperref}
\usepackage{hyperref}
\usepackage{color}

\begin{document}

%%%%%%%%%%%%%%%%%%%%%%%%%%%%%%%%%%%%%%%%%%%

%\newcommand{\gtrsim}{ \mathop{}_{\textstyle \sim}^{\textstyle >} }
%\newcommand{\lesssim}{ \mathop{}_{\textstyle \sim}^{\textstyle <} }
\newcommand{\vev}[1]{ \left\langle {#1} \right\rangle }
\newcommand{\bra}[1]{ \langle {#1} | }
\newcommand{\ket}[1]{ | {#1} \rangle }
\newcommand{\eV}{ \ {\rm eV} }
\newcommand{\KeV}{ \ {\rm keV} }
\newcommand{\MeV}{\  {\rm MeV} }
\newcommand{\GeV}{\  {\rm GeV} }
\newcommand{\TeV}{\  {\rm TeV} }
\newcommand{\1}{\mbox{1}\hspace{-0.25em}\mbox{l}}
\newcommand{\Red}[1]{{\color{red} {#1}}}

\newcommand{\lmk}{\left(}  
\newcommand{\rmk}{\right)}
\newcommand{\lkk}{\left[}  
\newcommand{\rkk}{\right]}
\newcommand{\lhk}{\left \{ }  
\newcommand{\rhk}{\right \} }
\newcommand{\del}{\partial}  
\newcommand{\la}{\left\langle} 
\newcommand{\ra}{\right\rangle}
\newcommand{\half}{\frac{1}{2}}

\newcommand{\bea}{\begin{array}}
\newcommand{\eea}{\end{array}}
\newcommand{\beq}{\begin{eqnarray}}
\newcommand{\eeq}{\end{eqnarray}}

\newcommand{\dd}{\mathrm{d}}
\newcommand{\Mpl}{M_{\rm Pl}}
\newcommand{\mg}{m_{3/2}}
\newcommand{\abs}[1]{\left\vert {#1} \right\vert}
\newcommand{\mphi}{m_{\phi}}
\newcommand{\Hz}{\ {\rm Hz}}
\newcommand{\for}{\quad \text{for }}
\newcommand{\Min}{\text{Min}}
\newcommand{\Max}{\text{Max}}
\newcommand{\Kahler}{K\"{a}hler }
\newcommand{\cphi}{\varphi}
\newcommand{\Tr}{\text{Tr}}
\newcommand{\diag}{{\rm diag}}

\newcommand{\SUf}{SU(3)_{\rm f}}
\newcommand{\Upq}{U(1)_{\rm PQ}}
\newcommand{\Zpq}{Z^{\rm PQ}_3}
\newcommand{\Cpq}{C_{\rm PQ}}
\newcommand{\ubar}{u^c}
\newcommand{\dbar}{d^c}
\newcommand{\ebar}{e^c}
\newcommand{\nubar}{\nu^c}
\newcommand{\Ndw}{N_{\rm DW}}
\newcommand{\Fpq}{F_{\rm PQ}}
\newcommand{\fpq}{v_{\rm PQ}}
\newcommand{\Br}{{\rm Br}}
\newcommand{\Lag}{\mathcal{L}}
\newcommand{\Lqcd}{\Lambda_{\rm QCD}}

\newcommand{\ji}{j_{\rm inf}} 
\newcommand{\jb}{j_{B-L}} 
\newcommand{\M}{M} 
\newcommand{\im}{{\rm Im} }
\newcommand{\re}{{\rm Re} }

%added by FT
\def\lrf#1#2{ \left(\frac{#1}{#2}\right)}
\def\lrfp#1#2#3{ \left(\frac{#1}{#2} \right)^{#3}}
\def\lrp#1#2{\left( #1 \right)^{#2}}
\def\REF#1{Ref.~\cite{#1}}
\def\SEC#1{Sec.~\ref{#1}}
\def\FIG#1{Fig.~\ref{#1}}
\def\EQ#1{Eq.~(\ref{#1})}
\def\EQS#1{Eqs.~(\ref{#1})}
\def\TEV#1{10^{#1}{\rm\,TeV}}
\def\GEV#1{10^{#1}{\rm\,GeV}}
\def\MEV#1{10^{#1}{\rm\,MeV}}
\def\KEV#1{10^{#1}{\rm\,keV}}
\def\blue#1{\textcolor{blue}{#1}}
\def\red#1{\textcolor{blue}{#1}}

\newcommand{\fa}{f_{a}}
\newcommand{\Uh}{U(1)$_{\rm H}$}
\newcommand{\osc}{_{\rm osc}}

\newcommand{\mav}{\left. m_a^2 \right\vert_{T=0}}
\newcommand{\mat}{m_{a, {\rm QCD}}^2 (T)}
\newcommand{\mam}{m_{a, {\rm M}}^2 }

%%%%%%%%%%%%%%%%%%%%%%%%%%%%%%%%%%%%%%%%%%%%%%%%%%%%%%%%%%%%%%%

\preprint{
TU-1011; \\
%IPMU 15-0xxx; \\
DESY 15-216
}

\title{
Suppressing the QCD Axion Abundance by Hidden Monopoles
}

\author{
Masahiro Kawasaki
}
\affiliation{Institute for Cosmic Ray Research, 
The University of Tokyo, 
Kashiwa, Chiba 277-8582, Japan}
\affiliation{Kavli IPMU (WPI), UTIAS, 
The University of Tokyo, 
Kashiwa, Chiba 277-8583, Japan}

\author{
Fuminobu Takahashi
}
\affiliation{Department of Physics, Tohoku University, 
Sendai, Miyagi 980-8578, Japan} 
\affiliation{Kavli IPMU (WPI), UTIAS, 
The University of Tokyo, 
Kashiwa, Chiba 277-8583, Japan}

\author{
Masaki Yamada
}
\affiliation{Institute for Cosmic Ray Research, 
The University of Tokyo, 
Kashiwa, Chiba 277-8582, Japan}
\affiliation{Kavli IPMU (WPI), UTIAS, 
The University of Tokyo, 
Kashiwa, Chiba 277-8583, Japan}
\affiliation{
Deutsches Elektronen-Synchrotron DESY, 
22607 Hamburg, Germany
}

\date{\today}

\begin{abstract} 
We study the Witten effect of hidden monopoles on the QCD axion dynamics, and show that its
abundance as well as isocurvature perturbations can be significantly suppressed if there is a sufficient 
amount of hidden monopoles. When the hidden monopoles make up a significant fraction of 
dark matter, the Witten effect suppresses the abundance of axion with the decay constant 
smaller than $10^{12} \GeV$. The cosmological domain wall problem of the QCD axion
can also be avoided, relaxing the upper bound on the decay constant 
when the Peccei-Quinn symmetry is spontaneously broken after inflation.
\end{abstract}

\maketitle

%%%%%%%%%%%%%%%%%%%%%%%%%%%%%%%%%%%%%%%%%%%%%%%%%%%%%%%%%%%%%%%%
\section{Introduction
\label{sec:introduction}}
%%%%%%%%%%%%%%%%%%%%%%%%%%%%%%%%%%%%%%%%%%%%%%%%%%%%%%%%%%%%%%%%

The smallness of the strong CP phase is an outstanding mystery in particle physics, and in particular,
it lacks any obvious anthropic explanation. The most natural solution is the Peccei-Quinn (PQ) mechanism, 
where an anomalous global symmetry is assumed to be broken spontaneously~\cite{Peccei:1977hh}. 
The associated pseudo-Nambu Goldstone boson, called the axion, 
obtains an effective mass via the QCD instanton effect~\cite{Weinberg:1977ma, 'tHooft:1976up}
and the CP phase is dynamically cancelled at the resulting potential minimum. 
Accordingly, the axion coherent oscillation
is necessarily produced by the misalignment mechanism 
and it contributes to dark matter (DM)~\cite{Preskill:1982cy} 
(See Refs.~\cite{Kim:2008hd,Ringwald:2012hr,Kawasaki:2013ae} for recent reviews). 
However, there are some problems in axion cosmology.

When the PQ symmetry is spontaneously broken before inflation, 
the abundance of the QCD axion dark matter depends on the axion decay constant $f_a$ and the initial misalignment angle $\theta_i$~\cite{Bae:2008ue}: 
\beq
\Omega_a h^2 \;\simeq\; 0.2\, \theta_i^2 \lrfp{f_a}{\GEV{12}}{1.19}, 
\label{Omegaa}
\eeq
where $h$ is the present-day Hubble parameter in units of $100$ km s$^{-1}$ Mpc$^{-1}$,
and anharmonic effects are neglected.
One can see  that the observed  DM abundance can be naturally 
explained by the coherent oscillations of the axion with $\theta_i = {\cal O}(1)$ and $f_a \simeq \GEV{12}$.
On the other hand, for $f_a \gg \GEV{12}$ as suggested by the string theory~\cite{Witten:1984dg, Svrcek:2006yi}, the initial misalignment
angle must be finely tuned as $\theta_i \ll 1$ to avoid the overclosure of the Universe. 
In addition, there is another problem related to the energy scale of inflation. 
Since the axion is massless during inflation, 
it acquires quantum fluctuations, $\delta a \simeq H_{\rm inf}/2\pi$, where
$H_{\rm inf}$ is the Hubble parameter during inflation~\cite{Axenides:1983hj}. 
As a result, the axion DM has isocurvature fluctuation,
which is constrained by observations, setting a tight upper bound on the inflation scale~\cite{Ade:2015lrj}.
In particular, there is a strong tension between the axion DM  and high-scale inflation. 
There have been proposed various ways to avoid or ameliorate the isocurvature limit on the axion DM;
restoration of the PQ symmetry~\cite{Linde:1990yj,Lyth:1992tw}, 
entropy dilution~\cite{Lazarides:1987zf, Dimopoulos:1988pw, Banks:1996ea}, 
time-dependent axion decay constant~\cite{Linde:1990yj,Linde:1991km,Kasuya:1996ns,Nakayama:2015pba,Harigaya:2015hha},
stronger QCD in the early Universe~\cite{Jeong:2013xta,Choi:2015zra} (see also Ref.~\cite{Dvali:1995ce}),
explicit breaking of the PQ symmetry~\cite{Dine:2004cq,Higaki:2014ooa,Dine:2014gba,Kawasaki:2015lea}, 
and a non-minimal kinetic term of the PQ scalar~\cite{Folkerts:2013tua}.

When the PQ symmetry is spontaneously broken after inflation, 
an axionic string and wall system appears at the QCD phase transition~\cite{Vilenkin:1982ks}. 
When the domain wall number is unity,  the string-wall network collapses due to the tension of domain walls. 
The axions are produced from those topological defects as well as the misalignment mechanism 
and the total abundance is given by~\cite{Kawasaki:2014sqa}
\beq
 \Omega_a h^2 \simeq 4 
 \lrfp{f_a}{\GEV{12}}{1.19}. 
 \label{eq2}
\eeq
On the other hand, 
when the domain wall number is greater than unity, 
the cosmic string and domain wall system is stable 
and soon dominates the Universe~\cite{Zeldovich:1974uw, Sikivie:1982qv}. 
The resulting Universe would be highly inhomogeneous, so that such scenario is excluded. 
This is known as the axion domain wall problem.

In this Letter we provide a novel solution to the above cosmological problems 
based on an axion coupling to monopoles of a hidden U(1)$_{\rm H}$ gauge symmetry.
In the presence of a CP violating  $\theta$-term,
monopoles acquire a non-zero electric charge and become dyons due to the Witten effect~\cite{Witten:1979ey}.
A non-zero  $\theta$ costs more energy as the mass of dyon is heavier than the monopole mass.
As a result, if $\theta$ is replaced with a dynamical axion field, 
the axion  acquires an extra potential.
 The Witten effect on the QCD axion was studied in Ref.~\cite{Fischler:1983sc}, where
monopoles were assumed to have an ordinary electromagnetic U(1)$_{\rm EM}$ charge. The
effect, however, turned out to be extremely small because of the tight observational constraints on the abundance
of the monopoles with U(1)$_{\rm EM}$ charge. Here we focus on hidden magnetic monopoles that are
much less constrained, and study their Witten effect  on the QCD axion dynamics, also taking account of the adiabatic suppression 
mechanism as well as various theoretical/cosmological possibilities.
We find that the axion abundance (and therefore the isocurvature perturbation) is suppressed efficiently 
and the domain wall problem is avoided. 
Instead of the axion,  the hidden monopoles (as well as vector fields) may make up the significant 
fraction of DM. Interestingly, they have non-negligible self-interactions, which could ameliorate 
small-scale tensions of the $\Lambda$CDM~\cite{Khoze:2014woa}.

%%%%%%%%%%%%%%%%%%%%%%%%%%%%%%%%%%%%%%%%%%%%%%%%%%%%%%%%%%%%%%%%
\section{The Witten effect on the QCD axion 
%dynamics
\label{sec:witten}}
%%%%%%%%%%%%%%%%%%%%%%%%%%%%%%%%%%%%%%%%%%%%%%%%%%%%%%%%%%%%%%%%

First, we summarize properties of the axion for later use.
The axion obtains an effective mass from QCD instanton effects~\cite{'tHooft:1976up}. 
At a temperature higher than $\Lqcd$, 
it is given by 
\beq
 \mat 
 \simeq 
 c_T \frac{\Lqcd^4}{\fa^2} 
 \lrfp{T}{ \Lqcd }{-n}, 
 \label{axion mass at T}
\eeq
where $c_T \simeq 1.68 \times 10^{-7}$, $n=6.68$, and $\Lqcd = 400 \MeV$~\cite{Kim:2008hd,Wantz:2009it}. 
The axion starts to oscillate around the CP conserving minimum 
at the temperature of 
\beq
 T_{\rm osc, 0} 
 &\simeq& 
 \Lqcd 
 \lrfp{90 c_T \Mpl^2}{\pi^2 g_* (T_{\rm osc, 0}) \fa^2}{1/(4+n)} 
 \\
 &\simeq& 2.8 \GeV \lrfp{\fa}{10^{10} \GeV}{-0.187}, 
\eeq
where $\Mpl$ ($\simeq 2.4 \times 10^{18} \GeV$) is the reduced Planck mass. 
The parameter $g_* (T)$ is the effective number of relativistic particles in the plasma 
and we use $g_* (T_{\rm osc, 0}) \approx 85$. 
For $T \lesssim100 \MeV$,
the axion mass  is approximately given by 
\beq
 \left. m_a \right\vert_{T=0} \simeq 
 \frac{z}{(1 + z)^2} \frac{m_\pi f_\pi}{\fa}, 
\eeq
where $z$ ($\simeq 0.56$) is the ratio of $u$- and $d$-quark masses, 
and $m_\pi$ ($\simeq 140 \MeV$) is the pion mass. 
The axion coherent oscillation is necessarily induced during the QCD phase transition,
and its abundance is given by Eq.~(\ref{Omegaa}).

Now let us explain the Witten effect in a hidden Abelian gauge theory~\cite{Witten:1979ey}.
Supposing that there are no charged particles,
the Lagrangian is given by 
\beq
\mathcal{L} = 
 -  \frac{1}{4} F_{\mu \nu} F^{\mu \nu} - \frac{e^2 \theta }{64 \pi^2}\epsilon_{\mu \nu \sigma \rho} 
 F^{\mu \nu} F^{\sigma \rho}, 
\eeq
where 
$e$ is the gauge coupling constant of the hidden gauge theory.
One of the Maxwell's equation is  given by 
\beq
 \del_\mu \lkk F_{\mu \nu} + \frac{e^2 \theta}{16 \pi^2}  \epsilon_{\mu \nu \sigma \rho} F^{\sigma \rho}
 \rkk 
 = 0, 
\eeq
so that the Gauss's law for the electric field is now modified as 
\beq
 \nabla \cdot {\bm E} + \frac{e^2}{8\pi^2} \nabla \cdot 
 \lmk \theta  {\bm B} \rmk 
 = 0, 
 \label{Gauss's law}
\eeq
where $E_i \equiv F_{0i}$ and $B_i \equiv -1/2\, \epsilon_{ijk} F^{jk}$.
Although ${\bm E}$ and ${\bm B}$ are of the hidden electric and magnetic fields,
we use them for notational simplicity. 
When we introduce a magnetic monopole with a magnetic charge $g$,
the Gauss's law for the magnetic field is given by $\nabla \cdot {\bm B} = g (n_{M^+}-n_{M^-})$, 
where $n_{M+} \,(n_{M^-})$ is the number density of (anti)monopoles. 
Then,  Eq.~(\ref{Gauss's law}) implies that the monopole carries also an electric charge $q$,
which is proportional to $\theta$.
In fact, the usual charge quantization condition, $q/e = n$, is extended to
\beq
 \frac{q}{e} + \frac{e g}{8 \pi^2} \theta = n, 
 \label{Witten effect}
\eeq
where $n$ is an integer. The periodicity of $\theta \to \theta + 2\pi$ can be seen if one substitutes 
a magnetic charge of the monopole, $g = 4\pi/e$.%
\footnote{
In our convention, 
half-integer electric charges are allowed. 
}
Thus, the monopole becomes a dyon due to the Witten effect. 
The mass of dyon mass is heavier than the monopole mass, and so, a nonzero $\theta$
costs more energy.

The Witten effect on the axion dynamics was studied by Fischler and Preskill~\cite{Fischler:1983sc},
where, instead of the $\theta$ parameter,  the axion is coupled to the gauge field as 
\beq
{\cal L}_\theta = - \frac{e^2  }{64 \pi^2} \frac{a}{f_a}\epsilon_{\mu \nu \sigma \rho} 
 F^{\mu \nu} F^{\sigma \rho}.
 \label{aFF}
\eeq
Here we have assumed that the domain wall number is unity for the above coupling. 
The potential energy of a single monopole was estimated to be
\beq
 V_M
 \approx 
 \beta \fa  \frac{a^2}{\fa^2}, 
 \\
  \beta 
 = \frac{\alpha}{32 \pi^2} \frac{1}{r_c \fa}, 
\eeq
where  $\alpha \equiv e^2 / 4\pi$, $a$ denotes the asymptotic field value of the axion,
and  $r_c$ is the radius of the monopole core. The origin of $a$ is chosen so that it
coincides with $\theta = 0$.
When we consider a 't Hooft-Polyakov  monopole, 
$r_c$ is 
the inverse of the mass of heavy gauge fields, $m_W$. 
As a result, 
the energy density of the axion ground state 
in a plasma with monopoles and antimonopoles  is given by $U = n_M V_0$, 
where $n_M = n_{M^+}+n_{M^-}$. Thus, 
the axion effectively obtains a mass of~%
\footnote{
We define $n_M$ as the sum of the number densities of monopoles and anti-monopoles, 
so that the axion mass squared is different from the one in Ref.~\cite{Fischler:1983sc} by a factor two. 
}
\beq
 \mam (T) = 2 \beta \frac{n_M (T)}{\fa}. 
 \label{FP effect}
\eeq
The monopole number density  $n_M$  will be evaluated in 
Sec.~\ref{sec:abundance}. 

%%%%%%%%%%%%%%%%%%%%%%%%%%%%%%%%%%%%%%%%%%%%%%%%%%%%%%%%%%%%%%%%
\section{
Axion dynamics 
\label{sec:dynamics}}
%%%%%%%%%%%%%%%%%%%%%%%%%%%%%%%%%%%%%%%%%%%%%%%%%%%%%%%%%%%%%%%%

Now let us consider the dynamics of axion 
in a plasma with monopoles. 
Once monopoles are produced in thermal plasma, 
its number density decreases as $R^{-3}$, where $R$ is the scale factor. 
This means that 
the ratio  $\mam / H^2$ increases with time during the radiation dominated era. 
The Witten effect becomes relevant for the axion dynamics when 
$\mam (T_{\rm osc, 1}) \simeq H^2 (T_{\rm osc, 1})$ is satisfied. 
Here, the temperature $T_{\rm osc, 1}$ is written as 
\begin{align}
T_{\rm osc, 1} 
& \simeq 
 Y_M 
 \frac{8 \beta \Mpl^2}{\fa}, \\
& \simeq 65 {\rm\,GeV} \alpha^2 \lrf{\Omega_M h^2}{0.12} \lrfp{f_a}{10^{12}\,{\rm GeV}}{-2}.
\end{align}
Hereafter, we focus on the case that 
the axion mass squared Eq.~(\ref{FP effect}) is much larger than the Hubble parameter 
squared at the time of $T = T_{\rm osc, 0}$, i.e. we consider the case of $T_{\rm osc, 0} \ll T_{\rm osc, 1}$.

Next we consider the cosmological history of axion. 
If the PQ symmetry is broken before inflation, 
the axion stays at a certain initial phase until the temperature decreases to $T_{\rm osc, 1}$ when
the axion starts to oscillate.
The resulting axion-number-to-entropy ratio is given by 
\beq
 \frac{n_a}{s} 
 &\simeq& 
 \frac{ H_{\rm osc, 1} \theta_{\rm ini}^2 \fa^2/2}{s (T_{\rm osc, 1})} 
 \\
 &\simeq& 
 \sqrt{\frac{45}{32 \pi^2 g_*}} \frac{\theta_{\rm ini}^2 \fa^2}{T_{\rm osc, 1} \Mpl}, 
 \label{Y_a}
\eeq
where $\theta_{\rm ini}$ is an initial phase of axion.
On the other hand, 
if the PQ symmetry is spontaneously broken after inflation, 
cosmic strings form after the phase transition. 
When the temperature decreases to $T_{\rm osc, 1}$, 
each cosmic string becomes attached by a domain wall 
due to the axion mass coming from the Witten effect. 
Then the cosmic strings and domain walls disappear soon due to the tension of the domain wall.%
\footnote{
This is not the case if the domain wall number in the interaction (\ref{aFF}) is
greater than unity.
Note that the domain wall number associated with the QCD instanton effects 
can be greater than unity as the axion potential from the QCD instanton is still negligible
at $T = T_{\rm osc,1}$.
}
As a result, 
the axion is produced from the decay of those topological defects 
and its abundance is expected be approximately given by Eq.~(\ref{Y_a})
with the replacement of $\theta_{\rm ini}^2 \to 20$, based on the axion 
abundances (\ref{Omegaa}) and (\ref{eq2}).
The subsequent evolution does not depend on whether the PQ symmetry is broken before or after inflation.

\begin{figure}[t]
\centering 
\includegraphics[width=.40\textwidth, bb=0 0 360 343]{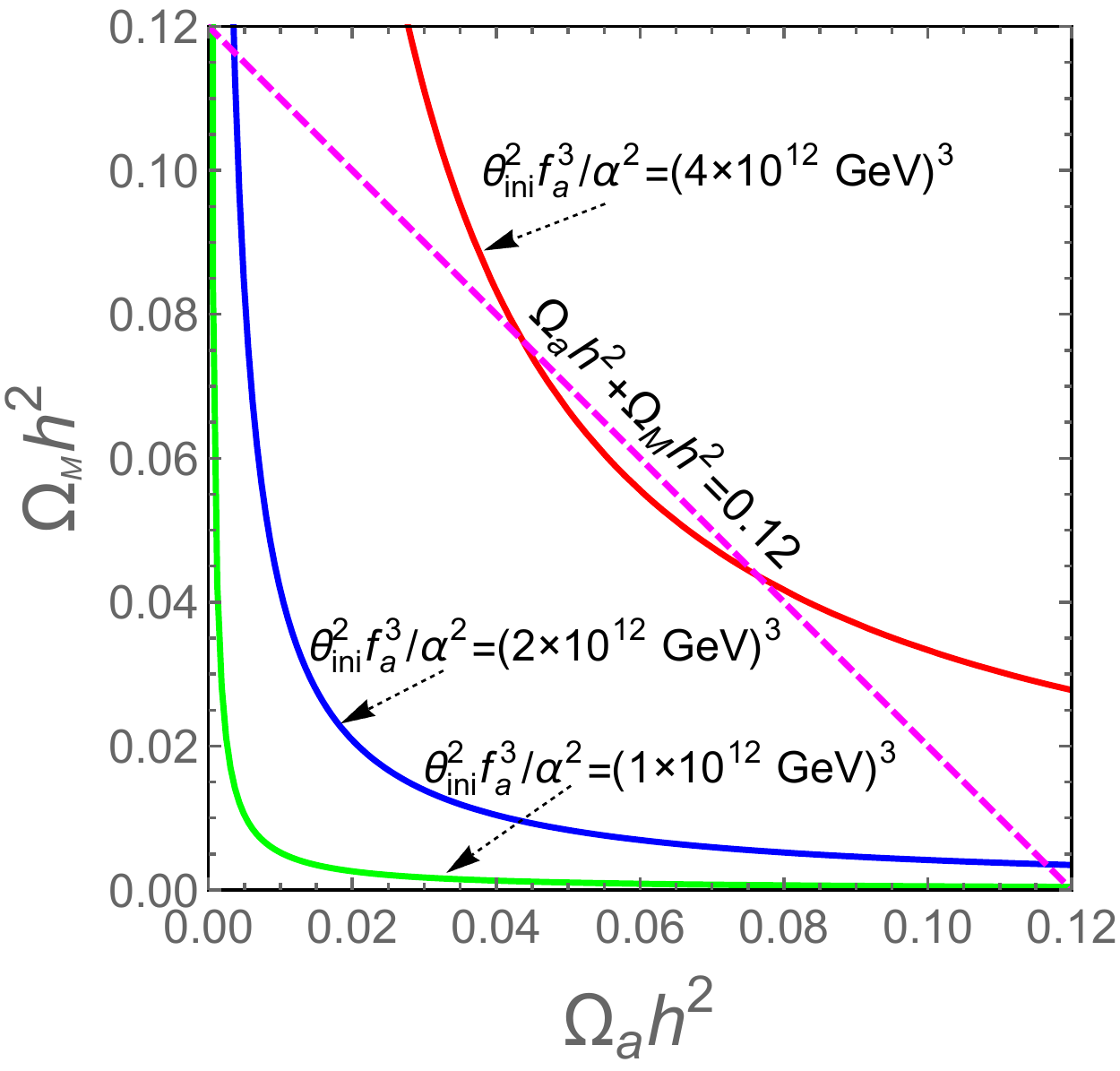} 
\caption{
Relation between the axion and monopole abundances.
We take $\theta_{\rm ini}^2 \fa^3 / \alpha^2 = (10^{12} \GeV)^3$ (green curve), 
$(2 \times 10^{12} \GeV)^3$ (blue curve), 
and $(4 \times 10^{12} \GeV)^3$ (red curve). 
The diagonal (magenta) dashed line represents the observed DM abundance. 
}
  \label{fig1}
\end{figure}

Around the time of QCD phase transition, 
the axion mass increases as Eq.~(\ref{axion mass at T}) due to 
QCD instanton effects. 
When $m_{a,{\rm QCD}}^2(T_{\rm osc, 2}) \simeq \mam (T_{\rm osc, 2})$, 
the potential minimum of axion changes adiabatically 
to the vacuum at which the strong CP phase is canceled if
$\mam  / H^2 \gtrsim 10^2$, i.e.,
\beq
f_a \lesssim 10^{12}\,{\rm GeV} \alpha^{1.1} \lrfp{\Omega_M h^2}{0.12}{0.55}. 
\eeq
Since the potential minimum changes adiabatically, 
the axion number density (in the comoving volume) is approximately conserved during this epoch 
and 
Eq.~(\ref{Y_a}) remains valid even after the QCD phase transition~\cite{Linde:1996cx, Takahashi:2015waa}. 
Let us emphasize that 
we do not have to assume any fine-tuning between the axion VEVs at the potential minima
due to the Witten effect and QCD instanton effect. 
We generically expect $\mathcal{O}(1)$ difference between the two minima,
but no extra axion oscillations are induced around the QCD phase transition 
due to the adiabatic suppression mechanism. 
The axion abundance is thus given by 
\begin{align}
 \Omega_a h^2 
& \simeq 
 3 \times 10^{-14} \theta_{\rm ini}^2 \frac{\fa}{T_{\rm osc, 1}} 
 \\
& \simeq 
3 \times 10^{-4}\, \frac{\theta_{\rm ini}^2}{\alpha^{2}} \lrf{0.12}{\Omega_M h^2} \lrfp{f_a}{\GEV{12}}{3} 
\label{Omegaa2}
\end{align}
where we use $g_* = 106.75$.  Interestingly, the axion abundance is inversely proportional
to that of the hidden monopoles. As long as U(1)$_{\rm H}$ is unbroken, the monopoles are stable and contribute
to DM. 
We plot Eq.~(\ref{Omegaa2}) in Fig.~\ref{fig1}, 
where we take $(\theta_{\rm ini}/\alpha)^{2/3} \fa= 10^{12} \GeV$ (green curve), 
$2 \times 10^{12} \GeV$ (blue curve), 
and $4 \times 10^{12} \GeV$ (red curve). 
The observed DM abundance can be explained at the intersection points of 
 each solid curve and the diagonal (magenta) dashed line representing $\Omega_{\rm DM} h^2 = 0.12$.
In order for the  abundance of axions and monopoles not to exceed 
the total DM abundance,  the axion decay constant is bounded from above:
\beq
 \fa \lesssim 4 \times 10^{12} \GeV 
 \lmk \frac{\alpha}{\theta_{\rm ini}} \rmk^{2/3}. 
 \label{fa bound}
\eeq

Since the axion abundance is related to the initial axion angle $\theta_{\rm ini}$, 
the quantum fluctuation of axion during inflation 
induces isocurvature modes to the CMB temperature fluctuations. 
The isocurvature constraint is written as~\cite{Ade:2015lrj} 
\beq
 \fa \gtrsim 3.4 \times 10^{4} H_{\rm inf} \frac{\Omega_a}{\Omega_{\rm DM}} \frac{1}{\theta_{\rm ini}}. 
\eeq
Thus,
the isocurvature constraint is weakened  if we can make the axion abundance smaller than the observed DM abundance.
The Witten effect indeed suppresses the axion abundance, thereby relaxing the isocurvature constraint
on the decay constant and the inflation scale.

%%%%%%%%%%%%%%%%%%%%%%%%%%%%%%%%%%%%%%%%%%%%%
%%%%%%%%%%%%%%%%%%%%%%%%%%%%%%%%%%%%%%%%%%%%%

%%%%%%%%%%%%%%%%%%%%%%%%%%%%%%%%%%%%%%%%%%%%%%%%%%%%%%%%%%%%%%%%
\section{Monopole abundance
\label{sec:abundance}}
%%%%%%%%%%%%%%%%%%%%%%%%%%%%%%%%%%%%%%%%%%%%%%%%%%%%%%%%%%%%%%%%
As an example,  let us consider  't Hooft-Polyakov  monopoles associated with the spontaneous breaking
of a hidden SU(2)$_{\rm H}$ gauge symmetry down to U(1)$_{\rm H}$ by the vacuum expectation  value of an 
adjoint Higgs $\la \phi_a \ra = (0,0,v)$, where $v$ is the symmetry breaking scale.
In this case,  there are charged massive gauge field $W^{\pm}$ of mass $m_W = e v$ and 
a massless hidden photon. The mass of magnetic monopole is given by $\sim 4 \pi v / e$. 
Assuming the second order phase transition, 
the monopole abundance is 
determined by the Kibble-Zurek mechanism: 
\beq
 Y_M \simeq 
 10^{-2} 
 \lrfp{30 T_c}{\sqrt{8 \pi} \Mpl}{3 \nu/(1+\nu)}, 
\eeq
where $\nu$ is a critical exponents~\cite{Murayama:2009nj}. 
At the tree level, $\nu = 1/2$, and it increases to $\nu \sim 0.7$ including
quantum corrections.

In the minimal set-up without any other light charged particles, 
the massive  gauge bosons $W^{\pm}$ are stable because they are the lightest particles
charged under the unbroken U(1)$_{\rm H}$. The abundances of the $W^\pm$ and monopoles 
were evaluated in Refs.~\cite{Baek:2013dwa, Khoze:2014woa}. 
In this case, the annihilation of monopoles 
sets an upper bound on its number density as~\cite{Khoze:2014woa}
\beq
 Y_M^{\rm max} 
 &\simeq& 
 \frac{2 \pi B}{g^2 x_*} \sqrt{\frac{45}{4 \pi^3 g_*}} \frac{m_M}{\Mpl} 
 \\
 B 
 &\equiv& 
 \frac{6 \zeta(3)}{\pi^2}, 
\eeq
where $x \equiv m_M / T$ and $x_* = \Min [x_{\rm f}, x_{\rm nr}]$. 
The parameter $x_{\rm f}$ is determined by the temperature at which 
the free streaming length of monopoles exceeds a capture radius of 
a monopole-anti-monopole bound state, 
while $x_{\rm nr}$ is determined by the temperature at which 
the massive gauge bosons become non-relativistic.%
\footnote{
In Ref.~\cite{Khoze:2014woa}, there is a typo in the last equality of (3.36), 
and as a result, their (3.37) should be multiplied with $\alpha^2$. 
}

The total abundance of the $W^\pm$ and monopoles is consistent with observations for e.g. 
$v \sim 10^{5} \GeV$ for $\alpha = {\cal O}(0.1)$, where the fraction of the monopole DM
is ${\cal O}(10)$\%.
Interestingly, both monopoles and $W$ have non-negligible self-interactions, and 
the small-scale tensions such as the `core-vs.-cusp'  and `too-big-to-fail' problems 
may be ameliorated~\cite{Baek:2013dwa,Khoze:2014woa}.

Finally, we comment on a kinetic mixing between the hidden 
and U(1)$_Y$ gauge bosons. 
Since we have introduced the adjoint Higgs field $\phi_a$ in the hidden sector,
we can write the following operator~\cite{Brummer:2009cs}: 
\beq
 \frac{\phi_a}{M} G_a^{\mu \nu} F_{\mu \nu} 
 \supset \frac{v}{M} G_3^{\mu \nu} F_{\mu \nu}, 
\eeq
where $G_a^{\mu \nu}$ and $F_{\mu \nu}$ are SU(2)$_{\rm H}$ and U(1)$_Y$ gauge fields, respectively. 
The parameter $M$ is a cutoff scale. 
This operator leads to the kinetic mixing of $v / M$, 
which is of order $10^{-13}$ for $v = 10^5 \GeV$ and $M = \Mpl$. 
Here we should note that 
the monopole in the hidden sector 
has an $\mathcal{O}(1)$ electric charge of hidden sector 
due to the Witten effect. 
This is because the axion stays at the minimum determined by the QCD instanton effect, 
which is generically deviated from the one determined by the Witten effect. 
Through the above kinetic mixing effect, 
the monopole as well as $W^\pm$ acquire a fractional SM electric charge 
of order $v / M$. While the current bound on the mini-charged DM is satisfied
for $v = 10^5 \GeV$ and $M= \Mpl$, the kinetic mixing may provide an interesting probe
of such DM candidates.

%%%%%%%%%%%%%%%%%%%%%%%%%%%%%%%%%%%%%%%%%%%%%%%%%%%%%%%%%%%%%%%%
\section{Discussion and conclusions
\label{sec:conclusion}}
%%%%%%%%%%%%%%%%%%%%%%%%%%%%%%%%%%%%%%%%%%%%%%%%%%%%%%%%%%%%%%%%

In this Letter,
we have proposed a novel mechanism to suppress the axion abundance based 
on the Witten effect.
If the QCD axion couples to a hidden Abelian gauge field,
the axion obtains a large effective mass in a plasma with monopoles. 
In particular we have focused on the case in which the Witten effect becomes important
before the QCD phase transition.

If the PQ symmetry is broken before inflation, 
the axion starts to oscillate earlier than usual, and the axion abundance is suppressed.
If the PQ symmetry is broken after inflation, there also appear axionic domain walls due to the
Witten effect. The domain walls are bounded by the cosmic strings associated with the spontaneous 
breaking of the PQ symmetry. As a result, the string-wall network soon collapses  due to the tension
of the domain walls. In either case, the axion abundance is determined 
when the effective mass becomes comparable to the Hubble parameter before the QCD phase transition.
Although the axion potential minimum is shifted to the strong CP conserving one
during the QCD phase transition, no extra coherent oscillations are induced as long as the effective mass
is much larger than the Hubble parameter. This is because
the axion follows its time-dependent minimum adiabatically so that its number density in the comoving
volume is conserved.  Thus, the final axion abundance can be significantly suppressed compared to
 the standard scenario.

We have found that the final axion abundance is inversely proportional to the monopole abundance (cf. Eq.~(\ref{Omegaa2})).
The monopoles are stable and 
its abundance is bounded above by the observed DM density. 
The suppression mechanism works when the axion decay constant is smaller than of order $10^{12} \GeV$,
if the DM is mainly composed of hidden monopoles. The monopoles (as well as the massive vector bosons
in the 't Hooft-Polyakov monopole) have self-interactions, which may relax small-scale problems of $\Lambda$CDM.

We have discussed the 't Hooft-Polyakov monopole as an explicit example. The observed DM abundance can be
explained by  both the monopoles and  massive gauge bosons for e.g. $v \sim 10^5$\,GeV and $\alpha = {\cal O}(0.1)$,
for which the monopoles occupy ${\cal O}(10)$\% of the total DM density~\cite{Baek:2013dwa, Khoze:2014woa}. 
It is also conceivable that, if the UV (electric) theory
becomes strongly coupled, light monopoles appear in the low-energy magnetic dual theory, and they may constitute
even larger fraction of the DM.

Finally we mention a possibility that the monopoles decay and disappear after the QCD phase transition. 
Suppose that the U(1)$_{\rm H}$ gauge symmetry is spontaneously broken by another Higgs field $\phi'$
at an energy scale of $v'$ ($1 \MeV \lesssim v' \lesssim 100 \MeV$). (For instance one may consider a doublet 
under SU(2)$_{\rm H}$.)
In this case, 
cosmic strings form at the phase transition 
and each monopole and anti-monopole pair is connected by 
a single cosmic string. 
This is known as the monopole confinement by the electron condensation in the Abelian gauge theory~\cite{Nambu:1974zg}. 
This implies that 
monopoles and anti-monopoles annihilate each other 
and disappear soon after the phase transition. The produced Higgs field may decay
into lighter SM particles through the portal coupling with the SM Higgs field, and such Higgs portal coupling 
may induce the invisible decay of the SM Higgs.
Thus we can avoid the upper bound on the monopole density.
Furthermore the massive gauge bosons also decay into the Higgs field $\phi'$. 
Therefore, there is no remnant in the hidden sector, which implies that the suppression 
of the axion abundance may be possible for $f_a \gg 10^{12}$\,GeV.
However, 
the  monopole abundance (before the spontaneous break down of U(1)$_{\rm H}$) 
may be modified because the energy dissipation rate of monopoles depend on the existence
of light charged particles.  We leave further detailed analysis of such scenario for future work.

\vspace{1cm}

%---------------SECTION------------------%
%
\section*{Acknowledgments}
M.Y. thanks W. Buchm\"{u}ller for kind hospitality at DESY. 
This work is supported by MEXT KAKENHI Grant Numbers 15H05889 (M.K. and F.T.) and 23104008 (F.T.),
JSPS KAKENHI Grant Numbers 25400248(M.K.), 24740135 (F.T.),  26247042(F.T.), and 26287039 (F.T.), 
JSPS Research Fellowships for Young Scientists (No.25.8715 (M.Y.)), 
World Premier International Research Center Initiative (WPI Initiative), MEXT, Japan,
and the Program for the Leading Graduate Schools, MEXT, Japan (M.Y.).
%
%---------------SECTION------------------%

\vspace{1cm}

%%%%%%%%%%%%%%%%%%%%%%%%%%%%%%%%%%%%

%%%%%%%%%%%%%%%%%%%%%%%%%%%%%%%%%%%%


\begin{thebibliography}{90}

  
  
%\cite{Peccei:1977hh}
\bibitem{Peccei:1977hh} 
  R.~D.~Peccei and H.~R.~Quinn,
  %``CP Conservation in the Presence of Instantons,''
  Phys.\ Rev.\ Lett.\  {\bf 38}, 1440 (1977); 
  %%CITATION = PRLTA,38,1440;%%
%
%\cite{Peccei:1977ur}
%\bibitem{Peccei:1977ur} 
%  R.~D.~Peccei and H.~R.~Quinn,
  %``Constraints Imposed by CP Conservation in the Presence of Instantons,''
  Phys.\ Rev.\ D {\bf 16}, 1791 (1977).
  %%CITATION = PHRVA,D16,1791;%%
  %1849 citations counted in INSPIRE as of 31 Mar 2015


  
\bibitem{Weinberg:1977ma} 
  S.~Weinberg,
  %``A New Light Boson?,''
  Phys.\ Rev.\ Lett.\  {\bf 40}, 223 (1978);
  %%CITATION = PRLTA,40,223;%%
%\cite{Wilczek:1977pj}
% \bibitem{Wilczek:1977pj} 
  F.~Wilczek,
  %``Problem of Strong p and t Invariance in the Presence of Instantons,''
  Phys.\ Rev.\ Lett.\  {\bf 40}, 279 (1978).
  %%CITATION = PRLTA,40,279;%%


  
  %\cite{'tHooft:1976up}
\bibitem{'tHooft:1976up} 
  G.~'t Hooft,
  %``Symmetry Breaking Through Bell-Jackiw Anomalies,''
  Phys.\ Rev.\ Lett.\  {\bf 37}, 8 (1976);
  %%CITATION = PRLTA,37,8;%%
  %2970 citations counted in INSPIRE as of 31 Mar 2015
%  
%  
  %\cite{'tHooft:1976fv}
%\bibitem{'tHooft:1976fv} 
%  G.~'t Hooft,
  %``Computation of the Quantum Effects Due to a Four-Dimensional Pseudoparticle,''
  Phys.\ Rev.\ D {\bf 14}, 3432 (1976)
  [Erratum-ibid.\ D {\bf 18}, 2199 (1978)].
  %%CITATION = PHRVA,D14,3432;%%
  %3413 citations counted in INSPIRE as of 31 Mar 2015
  
    



%\cite{Preskill:1982cy}
\bibitem{Preskill:1982cy} 
  J.~Preskill, M.~B.~Wise and F.~Wilczek,
  %``Cosmology of the Invisible Axion,''
  Phys.\ Lett.\ B {\bf 120}, 127 (1983); 
  %%CITATION = PHLTA,B120,127;%%
  %1030 citations counted in INSPIRE as of 31 Mar 2015
%  
  %\cite{Abbott:1982af}
%\bibitem{Abbott:1982af} 
  L.~F.~Abbott and P.~Sikivie,
  %``A Cosmological Bound on the Invisible Axion,''
  Phys.\ Lett.\ B {\bf 120}, 133 (1983);
  %%CITATION = PHLTA,B120,133;%%
  %995 citations counted in INSPIRE as of 31 Mar 2015
%  
  %\cite{Dine:1982ah}
%\bibitem{Dine:1982ah} 
  M.~Dine and W.~Fischler,
  %``The Not So Harmless Axion,''
  Phys.\ Lett.\ B {\bf 120}, 137 (1983).
  %%CITATION = PHLTA,B120,137;%%
  %980 citations counted in INSPIRE as of 31 Mar 2015


%\bibitem{QCD-axion}
  
  %\cite{Kim:2008hd}
\bibitem{Kim:2008hd} 
  J.~E.~Kim and G.~Carosi,
  %``Axions and the Strong CP Problem,''
  Rev.\ Mod.\ Phys.\  {\bf 82}, 557 (2010)
  [arXiv:0807.3125 [hep-ph]].
  %%CITATION = ARXIV:0807.3125;%%
  %284 citations counted in INSPIRE as of 13 juil. 2015
  
  %\cite{Ringwald:2012hr}
\bibitem{Ringwald:2012hr}
 A.~Ringwald,
 %``Exploring the Role of Axions and Other WISPs in the Dark Universe,''
 Phys.\ Dark Univ.\  {\bf 1} (2012) 116
 [arXiv:1210.5081 [hep-ph]].
 %%CITATION = ARXIV:1210.5081;%%

%\cite{Kawasaki:2013ae}
\bibitem{Kawasaki:2013ae} 
  M.~Kawasaki and K.~Nakayama,
  %``Axions: Theory and Cosmological Role,''
  Ann.\ Rev.\ Nucl.\ Part.\ Sci.\  {\bf 63}, 69 (2013)
  [arXiv:1301.1123 [hep-ph]].
  %%CITATION = ARXIV:1301.1123;%%
  %27 citations counted in INSPIRE as of 20 May 2014

  %\cite{Bae:2008ue}
\bibitem{Bae:2008ue} 
  K.~J.~Bae, J.~H.~Huh and J.~E.~Kim,
  %``Update of axion CDM energy,''
  JCAP {\bf 0809}, 005 (2008)
  [arXiv:0806.0497 [hep-ph]].
  %%CITATION = ARXIV:0806.0497;%%
  %71 citations counted in INSPIRE as of 14 Jul 2015

  
  
  %\cite{Witten:1984dg}
\bibitem{Witten:1984dg} 
  E.~Witten,
  %``Some Properties of O(32) Superstrings,''
  Phys.\ Lett.\ B {\bf 149}, 351 (1984).
  %%CITATION = PHLTA,B149,351;%%
  %639 citations counted in INSPIRE as of 10 Nov 2015
  
  
   %\cite{Svrcek:2006yi}
\bibitem{Svrcek:2006yi} 
  P.~Svrcek and E.~Witten,
  %``Axions In String Theory,''
  JHEP {\bf 0606}, 051 (2006)
  [hep-th/0605206].
  %%CITATION = HEP-TH/0605206;%%
  %321 citations counted in INSPIRE as of 11 juil. 2015 

    

  %\cite{Axenides:1983hj}
\bibitem{Axenides:1983hj} 
  M.~Axenides, R.~H.~Brandenberger and M.~S.~Turner,
  %``Development of Axion Perturbations in an Axion Dominated Universe,''
  Phys.\ Lett.\ B {\bf 126}, 178 (1983);
  %%CITATION = PHLTA,B126,178;%%
  %81 citations counted in INSPIRE as of 31 Mar 2015
%  
  %\cite{Seckel:1985tj}
%\bibitem{Seckel:1985tj} 
  D.~Seckel and M.~S.~Turner,
  %``Isothermal Density Perturbations in an Axion Dominated Inflationary Universe,''
  Phys.\ Rev.\ D {\bf 32}, 3178 (1985);
  %%CITATION = PHRVA,D32,3178;%%
  %123 citations counted in INSPIRE as of 31 Mar 2015
%  
  %\cite{Turner:1990uz}
%\bibitem{Turner:1990uz} 
  M.~S.~Turner and F.~Wilczek,
  %``Inflationary axion cosmology,''
  Phys.\ Rev.\ Lett.\  {\bf 66}, 5 (1991).
  %%CITATION = PRLTA,66,5;%%
  %108 citations counted in INSPIRE as of 31 Mar 2015


  %\cite{Ade:2015lrj}
\bibitem{Ade:2015lrj} 
  P.~A.~R.~Ade {\it et al.}  [Planck Collaboration],
  %``Planck 2015 results. XX. Constraints on inflation,''
  arXiv:1502.02114 [astro-ph.CO].
  %%CITATION = ARXIV:1502.02114;%%
  %148 citations counted in INSPIRE as of 04 juin 2015
    
    
    %\cite{Linde:1990yj}
\bibitem{Linde:1990yj} 
  A.~D.~Linde and D.~H.~Lyth,
  %``Axionic domain wall production during inflation,''
  Phys.\ Lett.\ B {\bf 246}, 353 (1990).
  %%CITATION = PHLTA,B246,353;%%
  %89 citations counted in INSPIRE as of 11 Nov 2015
  
  %\cite{Lyth:1992tw}
\bibitem{Lyth:1992tw} 
  D.~H.~Lyth and E.~D.~Stewart,
  %``Axions and inflation: String formation during inflation,''
  Phys.\ Rev.\ D {\bf 46}, 532 (1992).
  %%CITATION = PHRVA,D46,532;%%
  %60 citations counted in INSPIRE as of 11 Nov 2015
    
    
  %\cite{Lazarides:1987zf}
\bibitem{Lazarides:1987zf} 
  G.~Lazarides, C.~Panagiotakopoulos and Q.~Shafi,
  %``Relaxing the Cosmological Bound on Axions,''
  Phys.\ Lett.\ B {\bf 192}, 323 (1987).
  %%CITATION = PHLTA,B192,323;%%
  %57 citations counted in INSPIRE as of 11 Nov 2015
  
  %\cite{Dimopoulos:1988pw}
\bibitem{Dimopoulos:1988pw} 
  S.~Dimopoulos and L.~J.~Hall,
  %``Inflation and Invisible Axions,''
  Phys.\ Rev.\ Lett.\  {\bf 60}, 1899 (1988).
  %%CITATION = PRLTA,60,1899;%%
  %23 citations counted in INSPIRE as of 11 Nov 2015  
    
  %\cite{Banks:1996ea}
\bibitem{Banks:1996ea} 
  T.~Banks and M.~Dine,
  %``The Cosmology of string theoretic axions,''
  Nucl.\ Phys.\ B {\bf 505}, 445 (1997)
  [hep-th/9608197].
  %%CITATION = HEP-TH/9608197;%%
  %105 citations counted in INSPIRE as of 14 juil. 2015
  


%\cite{Linde:1991km}
\bibitem{Linde:1991km} 
  A.~D.~Linde,
  %``Axions in inflationary cosmology,''
  Phys.\ Lett.\ B {\bf 259}, 38 (1991).
  %%CITATION = PHLTA,B259,38;%%
  %608 citations counted in INSPIRE as of 04 juin 2015

  %\cite{Kasuya:1996ns}
\bibitem{Kasuya:1996ns} 
  S.~Kasuya, M.~Kawasaki and T.~Yanagida,
  %``Cosmological axion problem in chaotic inflationary universe,''
  Phys.\ Lett.\ B {\bf 409}, 94 (1997)
  [hep-ph/9608405];
  %%CITATION = HEP-PH/9608405;%%
  %12 citations counted in INSPIRE as of 17 Jun 2015
%  
%  
  %\cite{Kasuya:1997td}
%\bibitem{Kasuya:1997td} 
  S.~Kasuya, M.~Kawasaki and T.~Yanagida,
  %``Domain wall problem of axion and isocurvature fluctuations in chaotic inflation models,''
  Phys.\ Lett.\ B {\bf 415}, 117 (1997)
  [hep-ph/9709202]; 
  %%CITATION = HEP-PH/9709202;%%
  %11 citations counted in INSPIRE as of 17 juin 2015
 % 
  %
  %\cite{Kawasaki:2013iha}
%\bibitem{Kawasaki:2013iha} 
  M.~Kawasaki, T.~T.~Yanagida and K.~Yoshino,
  %``Domain wall and isocurvature perturbation problems in axion models,''
  JCAP {\bf 1311}, 030 (2013)
  [arXiv:1305.5338 [hep-ph]].
  %%CITATION = ARXIV:1305.5338;%%
  %7 citations counted in INSPIRE as of 17 juin 2015

  %\cite{Nakayama:2015pba}
\bibitem{Nakayama:2015pba} 
  K.~Nakayama and M.~Takimoto,
  %``Higgs inflation and suppression of axion isocurvature perturbation,''
  Phys.\ Lett.\ B {\bf 748}, 108 (2015)
  [arXiv:1505.02119 [hep-ph]].
  %%CITATION = ARXIV:1505.02119;%%
  %3 citations counted in INSPIRE as of 11 Nov 2015

%\cite{Harigaya:2015hha}
\bibitem{Harigaya:2015hha} 
  K.~Harigaya, M.~Ibe, M.~Kawasaki and T.~T.~Yanagida,
  %``Dynamics of Peccei-Quinn Breaking Field after Inflation and Axion Isocurvature Perturbations,''
  JCAP {\bf 1511}, no. 11, 003 (2015)
  [arXiv:1507.00119 [hep-ph]].
  %%CITATION = ARXIV:1507.00119;%%
  %2 citations counted in INSPIRE as of 11 Nov 2015
  
  
  %\cite{Dine:2004cq}
\bibitem{Dine:2004cq} 
  M.~Dine and A.~Anisimov,
  %``Is there a Peccei-Quinn phase transition?,''
  JCAP {\bf 0507}, 009 (2005)
  [hep-ph/0405256].
  %%CITATION = HEP-PH/0405256;%%
  %13 citations counted in INSPIRE as of 11 Jul 2015
  
   %\cite{Jeong:2013xta}
\bibitem{Jeong:2013xta} 
  K.~S.~Jeong and F.~Takahashi,
  %``Suppressing Isocurvature Perturbations of QCD Axion Dark Matter,''
  Phys.\ Lett.\ B {\bf 727}, 448 (2013)
  [arXiv:1304.8131 [hep-ph]];
  %%CITATION = ARXIV:1304.8131;%%
  %13 citations counted in INSPIRE as of 04 juin 2015  
%

%\cite{Choi:2015zra}
\bibitem{Choi:2015zra} 
  K.~Choi, E.~J.~Chun, S.~H.~Im and K.~S.~Jeong,
  %``Diluting the inflationary axion fluctuation by a stronger QCD in the early Universe,''
  Phys.\ Lett.\ B {\bf 750}, 26 (2015)
  [arXiv:1505.00306 [hep-ph]].
  %%CITATION = ARXIV:1505.00306;%%
  %4 citations counted in INSPIRE as of 11 Nov 2015
  
  %\cite{Dvali:1995ce}
\bibitem{Dvali:1995ce} 
  G.~R.~Dvali,
  %``Removing the cosmological bound on the axion scale,''
  hep-ph/9505253.
  %%CITATION = HEP-PH/9505253;%%
  %23 citations counted in INSPIRE as of 11 Nov 2015
  
  %\cite{Dine:2004cq}
\bibitem{Dine:2004cq} 
  M.~Dine and A.~Anisimov,
  %``Is there a Peccei-Quinn phase transition?,''
  JCAP {\bf 0507}, 009 (2005)
  [hep-ph/0405256].
  %%CITATION = HEP-PH/0405256;%%
  %14 citations counted in INSPIRE as of 10 Nov 2015
  
     %\cite{Higaki:2014ooa}
\bibitem{Higaki:2014ooa} 
  T.~Higaki, K.~S.~Jeong and F.~Takahashi,
  %``Solving the Tension between High-Scale Inflation and Axion Isocurvature Perturbations,''
  Phys.\ Lett.\ B {\bf 734}, 21 (2014)
  [arXiv:1403.4186 [hep-ph]];
  %%CITATION = ARXIV:1403.4186;%%
  %29 citations counted in INSPIRE as of 11 juil. 2015
  
      %\cite{Dine:2014gba}
%\cite{Dine:2014gba}
\bibitem{Dine:2014gba} 
  M.~Dine and L.~Stephenson-Haskins,
  %``Hybrid Inflation with Planck Scale Fields,''
  JHEP {\bf 1509}, 208 (2015)
  [arXiv:1408.0046 [hep-ph]].
  %%CITATION = ARXIV:1408.0046;%%
  %4 citations counted in INSPIRE as of 11 Nov 2015
  
  %\cite{Kawasaki:2015lea}
\bibitem{Kawasaki:2015lea} 
  M.~Kawasaki, M.~Yamada and T.~T.~Yanagida,
  %``Cosmologically safe QCD axion as a present from extra dimension,''
  Phys.\ Lett.\ B {\bf 750}, 12 (2015)
  [arXiv:1506.05214 [hep-ph]].
  %%CITATION = ARXIV:1506.05214;%%
  %3 citations counted in INSPIRE as of 11 Nov 2015

  %\cite{Folkerts:2013tua}
\bibitem{Folkerts:2013tua} 
  S.~Folkerts, C.~Germani and J.~Redondo,
  %``Axion Dark Matter and Planck favor non-minimal couplings to gravity,''
  Phys.\ Lett.\ B {\bf 728}, 532 (2014)
  [arXiv:1304.7270 [hep-ph]];
  %%CITATION = ARXIV:1304.7270;%%
  %16 citations counted in INSPIRE as of 04 juin 2015



    
  %\cite{Vilenkin:1982ks}
\bibitem{Vilenkin:1982ks} 
  A.~Vilenkin and A.~E.~Everett,
  %``Cosmic Strings and Domain Walls in Models with Goldstone and PseudoGoldstone Bosons,''
  Phys.\ Rev.\ Lett.\  {\bf 48}, 1867 (1982).
  %%CITATION = PRLTA,48,1867;%%
  %230 citations counted in INSPIRE as of 31 Mar 2015

%\cite{Kawasaki:2014sqa}
\bibitem{Kawasaki:2014sqa} 
  M.~Kawasaki, K.~Saikawa and T.~Sekiguchi,
  %``Axion dark matter from topological defects,''
  Phys.\ Rev.\ D {\bf 91}, no. 6, 065014 (2015)
  [arXiv:1412.0789 [hep-ph]].
  %%CITATION = ARXIV:1412.0789;%%
  %12 citations counted in INSPIRE as of 11 Nov 2015
  
    

  %\cite{Zeldovich:1974uw}
\bibitem{Zeldovich:1974uw} 
  Y.~B.~Zeldovich, I.~Y.~Kobzarev and L.~B.~Okun,
  %``Cosmological Consequences of the Spontaneous Breakdown of Discrete Symmetry,''
  Zh.\ Eksp.\ Teor.\ Fiz.\  {\bf 67}, 3 (1974)
  [Sov.\ Phys.\ JETP {\bf 40}, 1 (1974)].
  %%CITATION = ZETFA,67,3;%%
  %582 citations counted in INSPIRE as of 31 Mar 2015
  
  
  %\cite{Sikivie:1982qv}
\bibitem{Sikivie:1982qv} 
  P.~Sikivie [ADMX Collaboration],
  %``Of Axions, Domain Walls and the Early Universe,''
  Phys.\ Rev.\ Lett.\  {\bf 48}, 1156 (1982).
  %%CITATION = PRLTA,48,1156;%%
  %405 citations counted in INSPIRE as of 31 Mar 2015


%\cite{Witten:1979ey}
\bibitem{Witten:1979ey} 
  E.~Witten,
  %``Dyons of Charge e theta/2 pi,''
  Phys.\ Lett.\ B {\bf 86}, 283 (1979).
  %%CITATION = PHLTA,B86,283;%%
  %569 citations counted in INSPIRE as of 10 Nov 2015

%\cite{Fischler:1983sc}
\bibitem{Fischler:1983sc} 
  W.~Fischler and J.~Preskill,
  %``Dyon - Axion Dynamics,''
  Phys.\ Lett.\ B {\bf 125}, 165 (1983).
  %%CITATION = PHLTA,B125,165;%%
  %13 citations counted in INSPIRE as of 10 Nov 2015


  
  
  %\cite{Khoze:2014woa}
\bibitem{Khoze:2014woa} 
  V.~V.~Khoze and G.~Ro,
  %``Dark matter monopoles, vectors and photons,''
  JHEP {\bf 1410}, 61 (2014)
  [arXiv:1406.2291 [hep-ph]].
  %%CITATION = ARXIV:1406.2291;%%
  %14 citations counted in INSPIRE as of 10 Nov 2015
  


  
  

%\cite{Wantz:2009it}
\bibitem{Wantz:2009it} 
  O.~Wantz and E.~P.~S.~Shellard,
  %``Axion Cosmology Revisited,''
  Phys.\ Rev.\ D {\bf 82}, 123508 (2010)
  [arXiv:0910.1066 [astro-ph.CO]].
  %%CITATION = ARXIV:0910.1066;%%
  %79 citations counted in INSPIRE as of 31 Mar 2015
  
  

  
  
%\cite{Takahashi:2015waa}
\bibitem{Takahashi:2015waa} 
  F.~Takahashi and M.~Yamada,
  %``Strongly broken Peccei-Quinn symmetry in the early Universe,''
  JCAP {\bf 1510}, no. 10, 010 (2015)
  [arXiv:1507.06387 [hep-ph]].
  %%CITATION = ARXIV:1507.06387;%%
  %3 citations counted in INSPIRE as of 10 Nov 2015
  
  
  %\cite{Linde:1996cx}
\bibitem{Linde:1996cx} 
  A.~D.~Linde,
  %``Relaxing the cosmological moduli problem,''
  Phys.\ Rev.\ D {\bf 53}, 4129 (1996)
  [hep-th/9601083].
  %%CITATION = HEP-TH/9601083;%%
  %74 citations counted in INSPIRE as of 22 juin 2015






%\cite{Murayama:2009nj}
\bibitem{Murayama:2009nj} 
  H.~Murayama and J.~Shu,
  %``Topological Dark Matter,''
  Phys.\ Lett.\ B {\bf 686}, 162 (2010)
  [arXiv:0905.1720 [hep-ph]].
  %%CITATION = ARXIV:0905.1720;%%
  %24 citations counted in INSPIRE as of 10 Nov 2015

    %\cite{Baek:2013dwa}
\bibitem{Baek:2013dwa} 
  S.~Baek, P.~Ko and W.~I.~Park,
  %``Hidden sector monopole, vector dark matter and dark radiation with Higgs portal,''
  JCAP {\bf 1410}, no. 10, 067 (2014)
  [arXiv:1311.1035 [hep-ph]].
  %%CITATION = ARXIV:1311.1035;%%
  %24 citations counted in INSPIRE as of 11 Nov 2015


  %\cite{Brummer:2009cs}
\bibitem{Brummer:2009cs} 
  F.~Brummer and J.~Jaeckel,
  %``Minicharges and Magnetic Monopoles,''
  Phys.\ Lett.\ B {\bf 675}, 360 (2009)
  [arXiv:0902.3615 [hep-ph]]; 
  %%CITATION = ARXIV:0902.3615;%%
  %17 citations counted in INSPIRE as of 10 Nov 2015
%  
  %\cite{Bruemmer:2009ky}
%\bibitem{Bruemmer:2009ky} 
  F.~Brummer, J.~Jaeckel and V.~V.~Khoze,
  %``Magnetic Mixing: Electric Minicharges from Magnetic Monopoles,''
  JHEP {\bf 0906}, 037 (2009)
  [arXiv:0905.0633 [hep-ph]].
  %%CITATION = ARXIV:0905.0633;%%
  %30 citations counted in INSPIRE as of 10 Nov 2015
   
  
%\cite{Nambu:1974zg}
\bibitem{Nambu:1974zg} 
  Y.~Nambu,
  %``Strings, Monopoles and Gauge Fields,''
  Phys.\ Rev.\ D {\bf 10}, 4262 (1974); 
  %%CITATION = PHRVA,D10,4262;%%
  %803 citations counted in INSPIRE as of 05 Oct 2015
  G.~'t Hooft, in {\it High Energy Physics}, (Editorice Compositori, Bologna, 1975); 
  %
  %\cite{Mandelstam:1974pi}
%\bibitem{Mandelstam:1974pi} 
  S.~Mandelstam,
  %``Vortices and Quark Confinement in Nonabelian Gauge Theories,''
  Phys.\ Rept.\  {\bf 23}, 245 (1976).
  %%CITATION = PRPLC,23,245;%%
  %1014 citations counted in INSPIRE as of 18 Oct 2015

  
  
  
  
  
    
\end{thebibliography}
\end{document}